\documentclass[preprint, 3p]{elsarticle} 

\usepackage{graphicx}  
\usepackage{dcolumn}   
\usepackage{bm}        
\usepackage{amssymb}   
\usepackage[latin1]{inputenc}
\usepackage[english]{babel}
\usepackage{amsmath}
\usepackage{amssymb}
\usepackage{graphicx}
\usepackage{fancyhdr}
\usepackage{nomencl}
\usepackage{epigraph}
\usepackage{color}

\usepackage[pdfpagelabels,pdftex,final]{hyperref}
\begin{document}

\title{Point contact investigations of film and interface magnetoresistance of La$_{0.7}$Sr$_{0.3}$MnO$_3$ heterostructures on Nb:SrTiO$_{3}$}

\author[ntnup]{\AA smund Monsen} \address[ntnu]{Dept. of Physics, NTNU, 7491 Trondheim, Norway}
\author[ntnue]{Jos E. Boschker}\address[ntnue]{Dept. of Electronics and Telecommunications, NTNU, 7491 Trondheim, Norway}
\author[uu]{Per Nordblad} \address[uu]{Dept. of Engineering Sciences, Uppsala University, Box 534, SE-751 21 Uppsala, Sweden}
\author[uu]{Roland Mathieu}
\author[ntnue]{Thomas Tybell}
\author[ntnup]{Erik Wahlstr\"om \corref{cor1}}
\cortext[cor1]{Corresponding author: erik.wahlstrom@ntnu.no}




\begin{abstract}

STM based magnetotransport measurements of epitaxial La$_{0.7}$Sr$_{0.3}$MnO$_3$ 32 nm thick films with and without an internal LaMnO$_3$ layer (0-8 nm thick) grown on Nb doped SrTiO$_3$ are presented. The measurements reveal two types of low field magnetoresistance (LFMR) with a magnitude of $\sim 0.1-1.5\%$. One LFMR contribution is identified as a conventional grain boundary/domain wall scattering through the symmetric I-V characteristics, high dependence on tip placements and insensitivity to introduction of LaMnO$_3$ layers. The other contribution originates from the reverse biased Nb doped SrTiO$_3$ interface and the interface layer of La$_{0.7}$Sr$_{0.3}$MnO$_3$.  Both LFMR contributions display a field dependence indicative of a higher coercivity ($\sim$200 Oe) than the bulk film. LaMnO$_3$ layers are found to reduce the rectifying properties of the junctions, and sub micron lateral patterning by electron beam lithography enhances the diodic properties, in accordance with a proposed transport model based on the locality of the injected current.

\end{abstract}

\maketitle

\section{Introduction }
\label{sec:intro}

During the last decades mixed valence manganites have been widely studied, much owing to the discovery of the colossal magnetoresistive  effect (CMR) \cite{jin1994},  and half-metallicity \cite{park1998} in these compounds.  This in combination with their high Curie temperature has made them interesting candidates for spintronics applications \cite{mathews1997, sun1996, salamon2001, pantel2012, israel2007}, where the electron's spin degree of freedom is used as well as its charge \cite{wolf2001}. In particular there has been an effort put into understanding and designing high magnetoresistance (MR) all oxide tunnel junctions, often using  SrTiO$_3$ (STO) as barrier material \cite{sun1996, bowen2003, ogimoto2003, donnell2000}. 

Model systems in this aspect has been La$_{0.7}$Sr$_{0.3}$MnO$_3$ (LSMO)/(STO) heterostructures, which for Nb doped STO (Nb:STO, n-type) results in a Schottky barrier at the LSMO/STO interface \cite{ruotolo2007, susaki2007, postma2004}.

In LSMO, the main intrinsic MR effect is the CMR effect, dominating close to the Curie temperature, which requires substantial magnetic fields \cite{helmolt1993, jin1994}, and thus may be of limited device potential. In contrast to CMR, the extrinsic low field magneto resistance (LFMR) is mainly observed in the deep ferromagnetic regime, and thought to originate from grain boundaries (GB) and domain wall (DW) scattering. 

Grain boundary magnetoresistance (GBMR) have been investigated in numerous forms of LSMO;  polycrystalline bulk and thin films \cite{hwang1996, gupta1996}, powders \cite{coey1999b}, bi-crystal junctions \cite{evetts1998, mathur1997, mathieu2001b, mathieu2001c, gunnarson2004, gross2000}, step-edge junctions \cite{ziese1999}, and through laser patterning \cite{bibes1999}. There is a general consensus that spin polarized tunneling (SPT) between domains is responsible for the GBMR \cite{ ziese2002, siwach2008}, but mechanisms such as spin dependent scattering at GB are also suggested \cite{wang1998, gupta1996}. In contrast to samples containing natural and artificial GB, single crystal films  show no GBMR \cite{hwang1996, gupta1996, gross2000}.

The domain wall magnetoresistance (DWMR) in manganites have been proposed theoretically in the double exchange framework \cite{yamanaka1996, zhang1996} and claimed experimentally in confined sub-micron geometries \cite{mathur1999, arnal2007, ruotolo2007, wolfman2001}. However, the reported DWMR amplitude is larger than predicted, and it has been proposed that for the strongly electron correlated manganites, phase separation at the domain walls \cite{rzchowski2004, golosov2003, mathur2001} could be responsible.

For manganite-based spintronics  applications, the manganite-electrode interface requires understanding of the band bending and alignment, Schottky barrier heights and formation of interface states.  Efforts to resolve these questions on Nb:STO substrates have resulted in significant advances in the field. Characterization of highly rectifying junctions through current-voltage \cite{postma2004, susaki2007, xie2007}, capacitance-voltage \cite{ruotolo2007, luo2011}, photoemission spectroscopy \cite{minohora2007} and internal photoemission \cite{hikita2009} have resulted in a growing agreement on a thermally assisted tunneling transport mechanism in these structures, although contradictory results are reported \cite{sawa2005, xie2007}. 

Considerable MR have been reported in such junctions, with crossovers from negative to positive MR with temperature and bias current \cite{jin2005, lu2005, sun2004b, sheng2005}. It has been indicated that oxygen deficiencies and magnetocapacitance could be involved \cite{nakagawa2005}, but no consistent model yet exists for this intriguing phenomenon. However, the magnetoresistivity of such structures also includes the MR effects from the junction. Accordingly, there are several possible contributions to MR in the combined LSMO/STO system.

Here, we utilize a combination of local probing and bias dependent measurements enabled by scanning tunneling microscopy (STM) based point contact  measurements, to single out and address the dominant resistive contributions in epitaxial LSMO and LSMO/LaMnO$_3$/LSMO heterostructures on Nb:STO substrates.Through a combination of localized,  dynamic and static magnetoresistive measurements, and studying samples with and without inserted LMO layers we assess both the domain and interface contributions to the LFMR and assess their strength and coercivity.

\section{Experimental }
\label{sec:experimental}
The heterostructures were grown by pulsed laser deposition (PLD). A KrF laser ($\lambda =248$) with fluency of $\sim$ 2 J/cm$^2$ and a  repetition rate of 1 Hz was used to ablate stoichiometric  LSMO and LaMnO$_3$ (LMO) targets onto 10x10 mm$^2$ (001) oriented niobium doped (0.05 weight percent) STO substrates. (LSMO)$_{40}$/(LMO)$_n$/(LSMO)$_{40}$ heterostructures were grown without breaking vacuum, with n = 0 to 8 unit cells in even numbers.

Doped STO was chosen in order to have a conducting substrate with lattice parameters close to that of the pseudo cubic LSMO unit cell.
The as-received hydrofluoric (HF) acid-etched substrates were cleaned in acetone and ethanol before and after a 1 hour annealing step in flowing oxygen at 950$^{\circ}$ C to enhance step and terrace quality.

Before deposition the targets were pre-ablated for 5 minutes at 10 Hz to obtain a contamination free surface. During film growth the substrate temperature was held at 680$^{\circ}$ C and set at a distance of 45 mm from the target, while monitoring the growth with an $in$ $situ$ reflection high energy electron diffraction (RHEED) setup. The O$_2$ pressure was kept at 0.2 and 0.02 mbar for LSMO and LMO deposition, respectively. Reduced pressure was employed for the LMO layers as it is well known that LMO easily adopts excess oxygen from its stoichiometric phase \cite{roosmalen1994, huang1997, adamo2008}.

Magnetic characterization was performed utilizing a superconducting quantum interference device (SQUID) measurement system at temperatures from 15-370 K. The magnetic field was oriented in the film plane along the magnetically hard (100)$_{pc}$ and (010)$_{pc}$ directions \cite{steenbeck1999, tsui2000}. M-T curves were recorded with a background field of 50 Oe, after field cooling the samples from 370 to 15 K in 1000 Oe. 

Prior to further study, the films were cut in two pieces using a diamond dicing saw. One for reference, the other to be patterned laterally. A pattern of squares and ellipses with short axis ranging from 500-150 nm, oriented in the (100)$_{pc}$ direction, was chosen to confine the conduction paths, and to define the shape anisotropy for multi and single domain states. The lithographic steps are as follows; first a 60 nm thick layer of amorphous carbon is evaporated onto the sample in an electron beam evaporator at a base pressure of 10$^{-8}$ mbar. Then a 100 nm thick PMMA layer is applied, enough to achieve the undercut necessary for liftoff, and after electron beam exposure and development, a 20 nm chromium layer is evaporated onto the stack. Following liftoff in acetone, the chromium pattern is used as a mask for the carbon film in a directional oxygen plasma etcher (20 watt for 5 minutes), resulting in a Cr/C mask. This pattern is then finally transferred to the underlying thin film using accelerated Ar ions in an Oxford CAIBE with 500 volts acceleration, 20 mA ion current and at a rotating sample stage tilted 5 degrees off-normal. A final oxygen plasma step removes the residual carbon layer. 

For point contact characterization, the samples were loaded in a custom built STM, optimized for magnetic manipulations and point contact spectroscopy (PCS) \cite{saxegaard2010}. Measurements were conducted from 150 K to room temperature, with a rotatable magnetic field up to 800 Oe in a dry nitrogen atmosphere, and a sampling rate of 100 kHz. The field was applied parallel to the film surface. A modulation of 0.005 mA at 10 KHz  was added to the measurement signal in order to extract the dynamic response. Currents from 0.05 to 0.8 mA of both polarities were applied between the film surface and substrate, and both current and magnetic field were swept at rates from 1 Hz to DC. Magnetoresistance is defined as $MR=[R(H)-R(0)]/R(0)$.

STM tips were  made by standard chemical etching of copper and tungsten wires \cite{ekvall1999, ibe1990}, and mechanical clipping of platinum-iridium wires.  We reproduced all the characteristics described below using all three tip materials. In order to make contact to the sample, the STM tip was lowered onto the sample surface until a preset resistance was reached, or the piezo-tube was fully extended. A main experimental obstacle was to make contact to the LSMO surface. The oxide layer found on our samples after annealing/processing \cite{Monsen20121360} is not metallic and this in combination with the material hardness resulted in deformation of tips before the contact was formed. Although this prevented a detailed study of the position dependence of the magnetoresistive properties, we observed only slight deterioration of the surface layer upon tip indentation and the MR active parts of the samples remain intact after contacting the samples, even after injection of up to 0.8 mA through the point contact.

\begin{figure}[t]
\begin{center}
\includegraphics[width=8 cm]{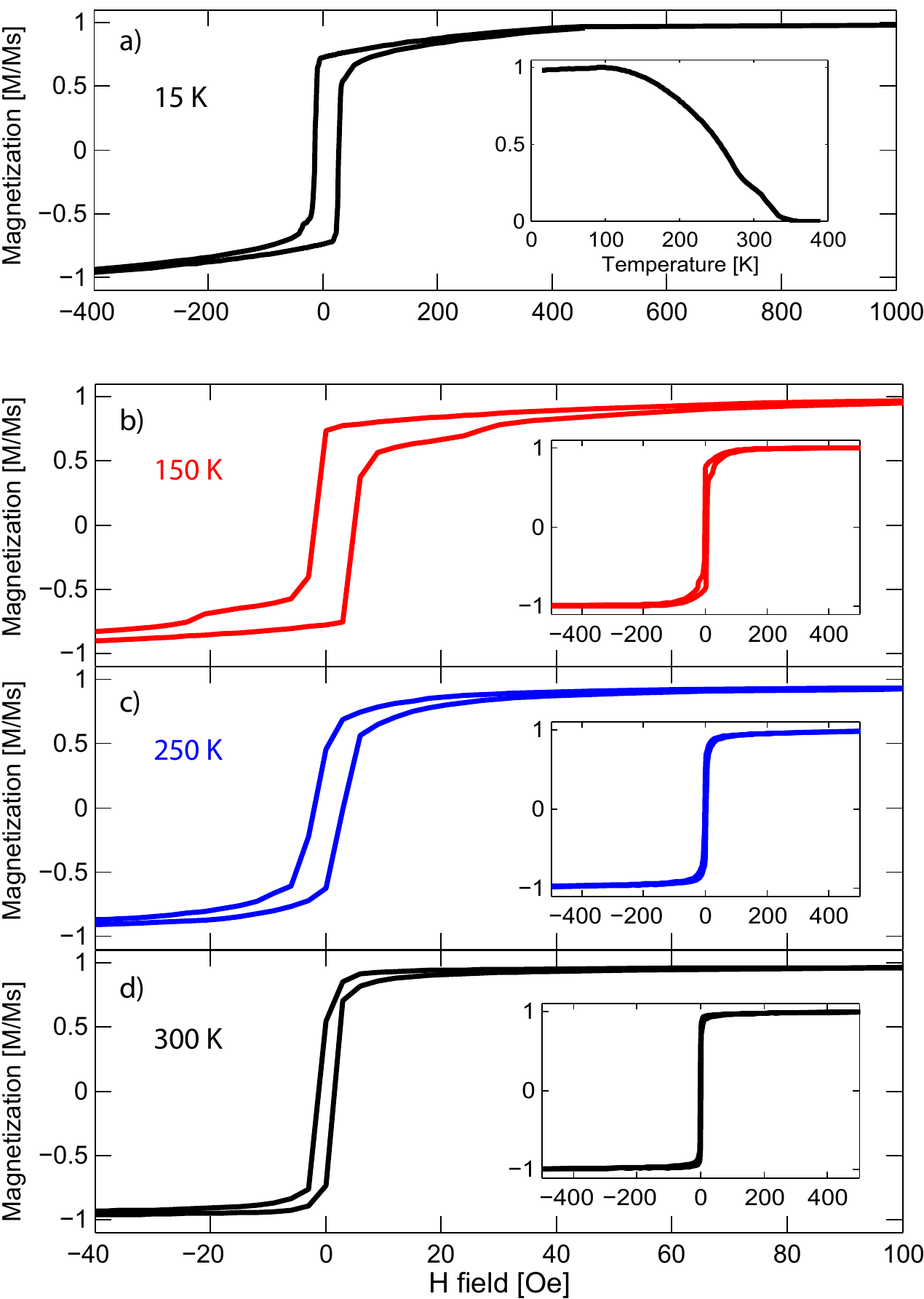}
\caption{\label{fig:SQUID} Normalized magnetization data for the heterostructure sample with 4 LMO layers. a-d): M-H curves at various temperatures. Insert in a) shows the M-T dependence, and inserts in (b-d) full field saturation M-H curves.  }
\end{center}
\end{figure}

\section{Results and discussion }
\label{sec:results}

\subsection{Structural and magnetic properties}

From the RHEED oscillations we concluded a layer by layer growth mode, confirmed by the 2D RHEED pattern recorded after ended deposition. AFM micrographs indicate the step and terrace structure of the substrates resulting from the miscut angle of $\sim$ 0.1 degrees to the (001) plane. 

The out of plane lattice parameters were found to be  $\sim$3.86 \AA \, as determined by fitting to  $\theta - 2\theta$ scans of the (001)$_{pc}$ and (002)$_{pc}$ Bragg peaks, comparable to that of single epitaxial LSMO thin films on STO substrates \cite{boschker2011}. The mosaic spread of the films was measured at the full width half maximum (FWHM), $\sim$ 0.025 degrees, comparable to that of the substrates, while the reciprocal space maps around the (103)$_{pc}$ peak of the films reveal an in-plane coherently strained structure. 

We deduce the Curie temperature ($T_c$) of the heterostructures to be $\sim$340 K, close to reported values for single LSMO films grown on STO \cite{boschker2011}, see insert in figure \ref{fig:SQUID} a). The small kink-like feature at around 100 K is present in all samples and tentatively ascribed to the structural transition of STO at 105 K \cite{cao2000, yamada1969}.

From the M-H curves at 15 K we determine the coercivities to be in the range 20 to 30 Oe, while saturation is reached at $\sim$ 500 Oe, yielding a magnetization of  [3.8 $\pm 0.07] \mu_{B}$  per Mn ion. At 150 K the coercivity is reduced to $\sim$  5 Oe, and  $\sim$ 1-2 Oe at 300 K.

\subsection{Point contact transport measurements}

\begin{figure}[t]
\begin{center}
\includegraphics[width=8 cm]{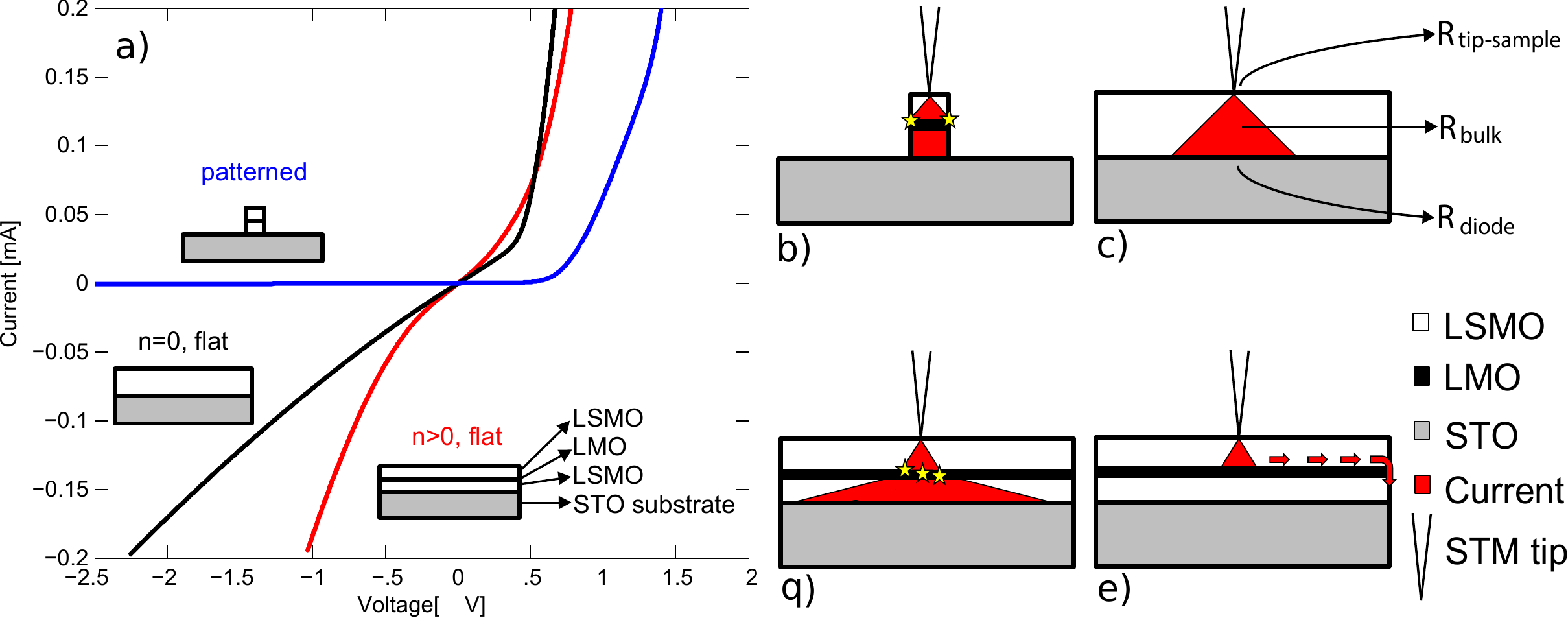}
\caption{\label{fig:pcs_iv_all}a) I-V curves at 150 K for patterned sample (blue), flat n=0 sample (black) and flat $>$0 sample (black), including illustrations of the various samples. The proposed current paths depending on sample. b) the patterned sample laterally confines the current, c) the flat film without LMO (n=0) layers yields a cone shaped injection. d) and e); the LMO inter-layers (n$>$0) act as heavy scatterers, and results in either larger interface area of probing, or alternate current path along possible edge defects.}
\end{center}
\end{figure}

Representative I-V curves from the point contact measurements, shown in figure \ref{fig:pcs_iv_all}, 
illustrate a classification in three groups, (\textit{i, ii, iii}), based on their diodic behavior;\\

\emph{(i, patterned samples, blue curve)} The laterally patterned samples all yielded clear diodic I-V characteristics, with no indication of breakdown with reverse bias for the voltage range inspected. No dependence on LMO thickness was detected. Such diodic behavior correlates well with reports on patterned junctions of LSMO and n-doped STO \cite{ruotolo2007, susaki2007,  postma2004}. \\ 

\emph{(ii, unpatterened sample, black curve)} The flat n=0 sample (LSMO/Nb:STO) displayed reduced diodic character with considerable transport also with reversed bias. This change in diodic behavior compared to the patterned samples indicates probing of a significantly larger film-substrate interface area. As such, local defects leading to tunneling in both forward and reverse bias become more prominent.  In addition to statistical appearance of defects, it has been shown that both LSMO and  STO interfaces display altered layers \cite{huijben2008, kim2010, stengel2006} which also could be laterally inhomogeneous \cite{kourkoutis2010}.
Although nonlinear I-V characteristics of the diodic barrier make calculation of the probed interface area  non-trivial, we expect a high density injected current just below contact, as  illustrated in figure \ref{fig:pcs_iv_all} c). Here the current follows a roughly 120 degree cone-shaped profile \cite{holm}, with a widening factor depending on local scattering, thus providing a limit for the LSMO/Nb:STO interface area probed.

\emph{(iii, sample with LMO layers inserted, red curve)} The flat n$>0$ samples displayed symmetrical I-V curves with little indication of diodicity. No dependence on LMO inter-layer thickness (n=2,4,6 or 8) were observed, and consequently no differentiation between the n$>$0 samples are made in the following.
This symmetric behavior, can be explained by expanding the above current spreading model, introducing the LMO inter-layers acting as additional scatterers (figure \ref{fig:pcs_iv_all} d-e).  As no significant dependence on the LMO inter-layer thickness is seen in the qualitative behavior of the I-V curves, we interpret the current cone as fully spread, i.e. probing a largely increased LSMO/Nb:STO interface. Similar symmetric curves are seen when using macro-contacts on unpatterned interfaces \cite{chen2007, chen2009}. Conversely, the transport in the patterned samples (\textit{i}) is less affected by scattering layers, as the current is laterally confined within the structure, limiting the probed interface area (figure \ref{fig:pcs_iv_all} b). 

\begin{figure}[t]
\begin{center}
\includegraphics[width=8.2 cm]{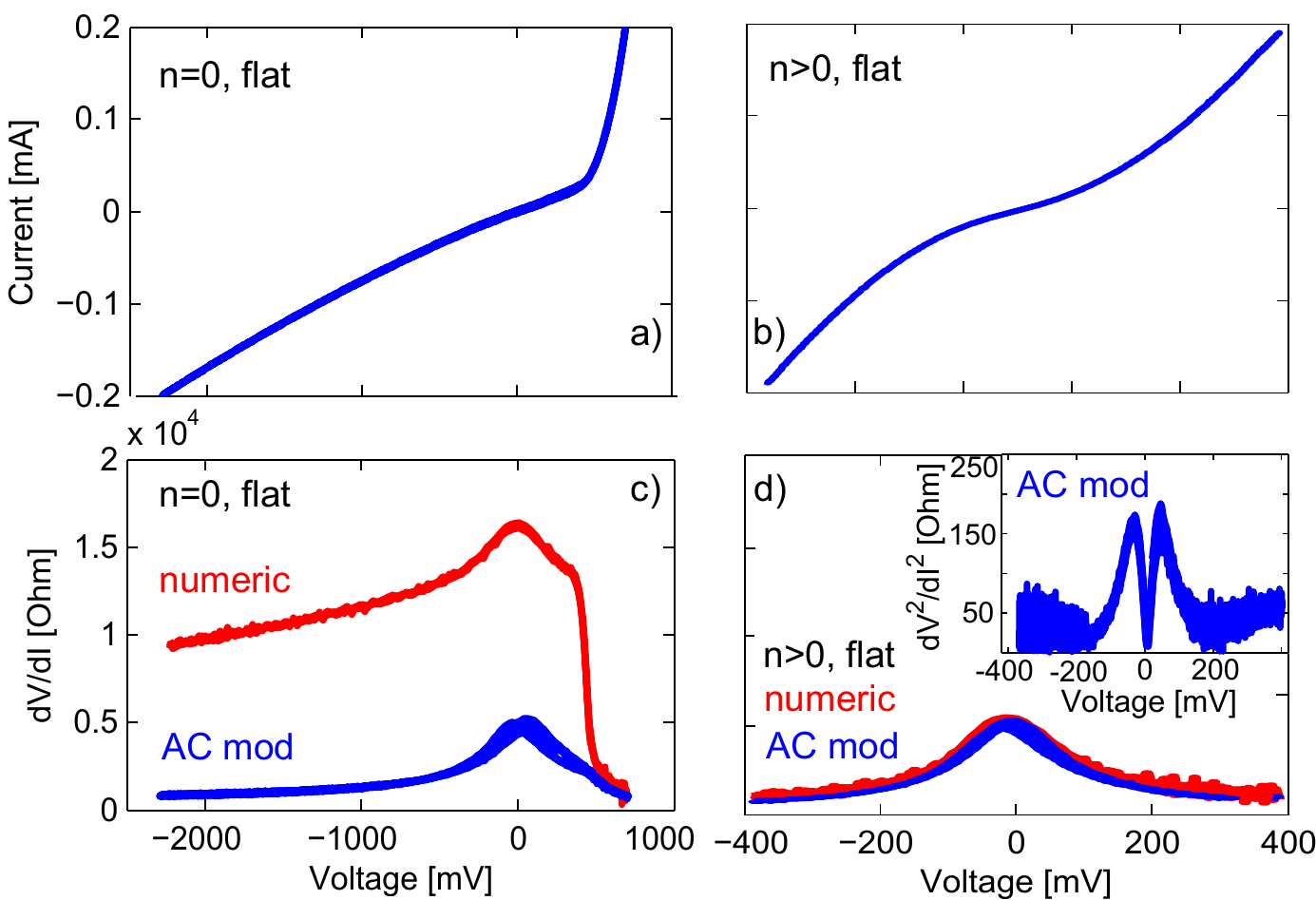}
\caption{\label{fig:pcs_dynR}  Dynamic resistance dependence on diodicity at 150 K. a)-b) I-V curves from samples displaying decreasing diodicity from left to right (n=0 and n$>$0). c)-d)  dynamic resistance from the same contacts; blue plots represent the dV/dI obtained from the AC-modulation of the probing current, red plots the numerical derived dV/dI. Insert in d) shows the double derivative derived from AC modulation. }
\end{center}
\end{figure}

Utilising samples with buried LMO layers accordingly allows for studying mainly the local contributions from the thin film point contact part of the system. This interpretation is corroborated by analyzing the dynamic response of the point contact; the diodic interface has a large junction capacitance associated with it, identified by a time-lag of the I-V sweeps (not shown). As the high frequency AC modulation of the current (added to obtain the dynamic response) bypasses this capacitance, we can selectively probe the system with and without the diodic contribution to the resistance, as shown in figure  \ref{fig:pcs_dynR}. Here we compare samples of varying diodic nature at 150 K, and their dynamic resistance both measured with AC modulation, and numerically derived from the I-V curves. As can be seen, the underlying resistive behavior is non ohmic; it decreases with increased current and is symmetric around zero bias voltage, although the DC bias voltage mapping is distorted by the diode junction. When we correct for this all samples display similar non-Ohmic dependence (not shown). This type of resistive dependence on current is often reported in manganite transport studies and is normally interpreted as electro-resistance (ER) \cite{ sun2007,  asamitsu1997, zhao2005}. Another possible cause is the often inferred  pseudo-gap  in LSMO \cite{mitra2003, singh2008}. This has been associated with polaronic states, where a strong electron-phonon coupling  trap charges in shallow localized states. The maxima in the double derivatives at $\sim \pm 80$ meV (insert figure \ref{fig:pcs_dynR} d), coincide with the maximum in the electron-phonon coupling as reported by point contact spectroscopy \cite{Mitra2002jj}, supporting this interpretation.

The observed diodic behavior of the structures is well documented for metallic manganite and n-doped STO interfaces \cite{ruotolo2007, susaki2007, zhang2002, postma2004, sawa2005}, even though conflicting reports regarding the exact transport mechanism exists \cite{sawa2005, susaki2007, xie2007}. We extract the Schottchy barrier parameters for patterned samples following the same procedures, and find that they correlate within error margin with what has been reported earlier \cite{ruotolo2007, susaki2007}.

\subsection{Point contact MR transport}

\begin{figure}[t]
\begin{center}
\includegraphics[width=7.5 cm]{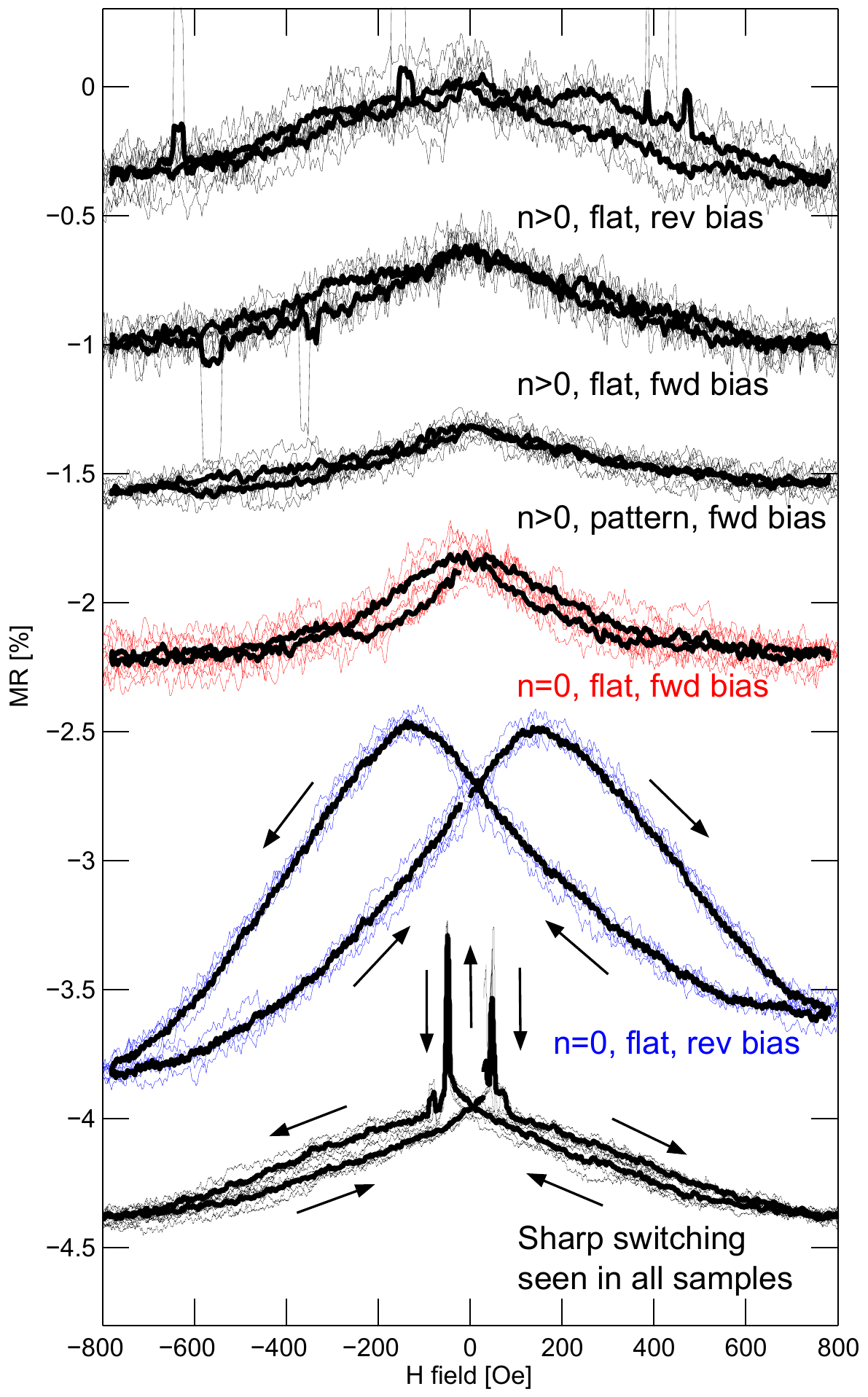}
\caption{\label{fig:MR_examples} Example MR curves of both forward and reverse bias at 0.1 mamp current for all sample groups at 150 K. Each plot consists of 5 periods of the B field sweep illustrating the stability of the effect, with the average superimposed. All plots except the top one are shifted downwards for clarity. The arrows give the direction of the sweep. The n=0 sample with diodic character diverges from the rest in the case of reverse bias. Bottom curve shows example of TMR like switching, infrequently appearing in all measurements. MR from the patterned sample was only ascertained under forward bias due to highly rectifying character. }
\end{center}
\end{figure}

All samples show negative MR in $\sim$ 2/3 of the point contacts, with a clear dependence on diodic character, as shown in figure \ref{fig:MR_examples}; The main point to be inferred, is that the MR resulting from reverse biasing the flat n=0 diodic sample (blue plot) is significantly larger than for forward biasing (red plot), whereas the MR of the non diodic samples with LMO inter-layers (n$>0$) exhibits MR independent of bias voltage polarity.

\begin{figure}[t]
\begin{center}
\includegraphics[width=7.5 cm]{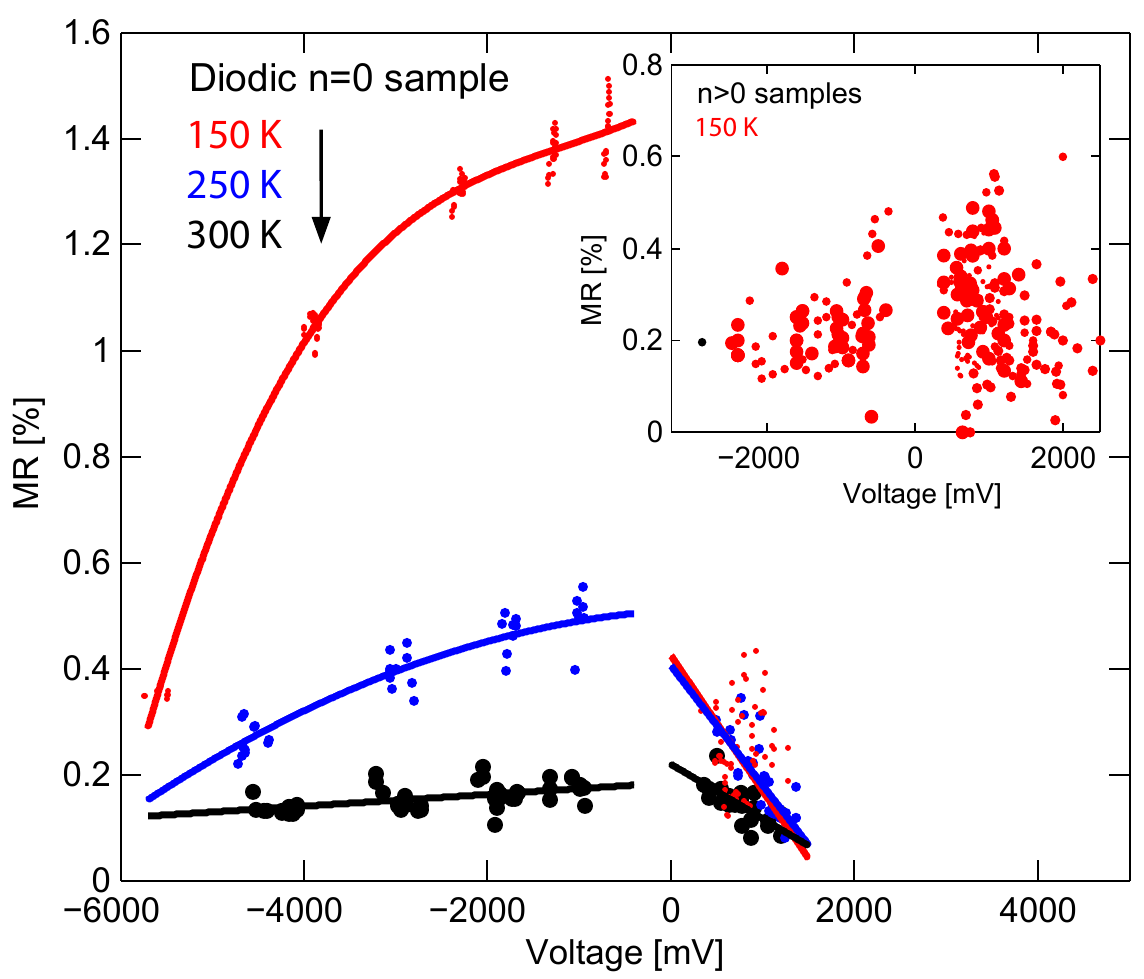}
\caption{\label{fig:MR_norm_summary} Summarized MR dependence on both bias and temperature for the diodic n=0 sample, summarized with statistics from a large number of point contacts and tips. Insert shows MR for the nondiodic n$>$0 samples and the patterned sample under forward bias. We refer back to figures \ref{fig:pcs_iv_all}  for I-V characteristics. All MR values are determined through comparison with the resistance at 800 Oe. }
\end{center}
\end{figure}

A more comprehensive summary of the MR, based on a large number of tips and contacts for all samples at varying temperatures and currents is shown in figure \ref{fig:MR_norm_summary}. Included in the insert is MR data from all n$>$0 samples; MR from the patterned samples is only plotted for forward bias as the highly rectifying properties prevented measurements under opposite voltage polarity. As no MR dependence on LMO inter-layer thickness is measured, the n$>$0 films are treated as one for the remainder of the paper.

We interpret the high and low amplitude MR to originate from two different mechanisms; the former stemming from an LSMO/Nb:STO interface effect related to the diodicity, and the latter from a more conventional GBMR \cite{xiao1992, hwang1996, gupta1996} or DWMR \cite{mathur1999, arnal2007, ruotolo2007, wolfman2001}. 

We base our assignment of two separate contributions to the MR on three observations. Firstly, the significant change in MR magnitude from 150 K to 250 K under reverse bias is non-existent under forward bias  (figure \ref{fig:MR_norm_summary}). The I-V characteristics for the flat n=0 sample correlate well with this temperature dependence, that is, the observed change in the ln(I) data (not shown) from 250 K to 150 K through a transition from a wide threshold voltage into a sharper well defined diode threshold voltage at 150 K. Secondly, the relative spread of datapoints in the high MR case is small compared to the scattered points in the low MR (Fig. \ref{fig:MR_norm_summary}). This corroborates the GB/DW explanation as the proximity to such boundaries will affect the MR value and thus greatly depends on the positioning of the STM tip during measurement. As each contact is done at a more or less random location, a stochastic spread of MR value is expected. Lastly, accompanying the high MR is a substantial magnetic hysteresis of ($\sim$ 170 Oe)  which is absent at 300 K. The low MR under forward bias shows little sign of this hysteresis (figure \ref{fig:MR_examples})

It is important to note that there is an energy dissipation in the system due to the relatively large probe current, and that it may lead to local heating of the films, predominantly in volumes with high energy dissipation close to the point contact. The low Curie temperature allows for us to access how large these effects are in the MR active regions studies here. For both MR effects discussed here, they are observable with equal probability at the highest temperature, which suggest that heating within these regions is confined to below 40 K. This can be expected since the MR effects are associated with interfaces/domain boundaries located quite far from the point contact.

\subsubsection*{Low amplitude MR}

The summarized low amplitude MR  (figure \ref{fig:MR_norm_summary}) displays a consistent amplitude from 150 to 250 K with a sharp drop at 300 K, and is in accordance with the GB observations for nanometer sized grains \cite{dey2005}. From the CMR effect normally dominating in single crystals and epitaxial films without GBs \cite{hwang1996, gross2000, gupta1996, mathieu2000}, the opposite temperature dependence is expected as it is connected to the magnetic phase transition and has a maximum at the Curie temperature. Thus we disregard CMR as the main contributor to the measured MR.

\begin{figure}[t]
\begin{center}
\includegraphics[width=7.5 cm]{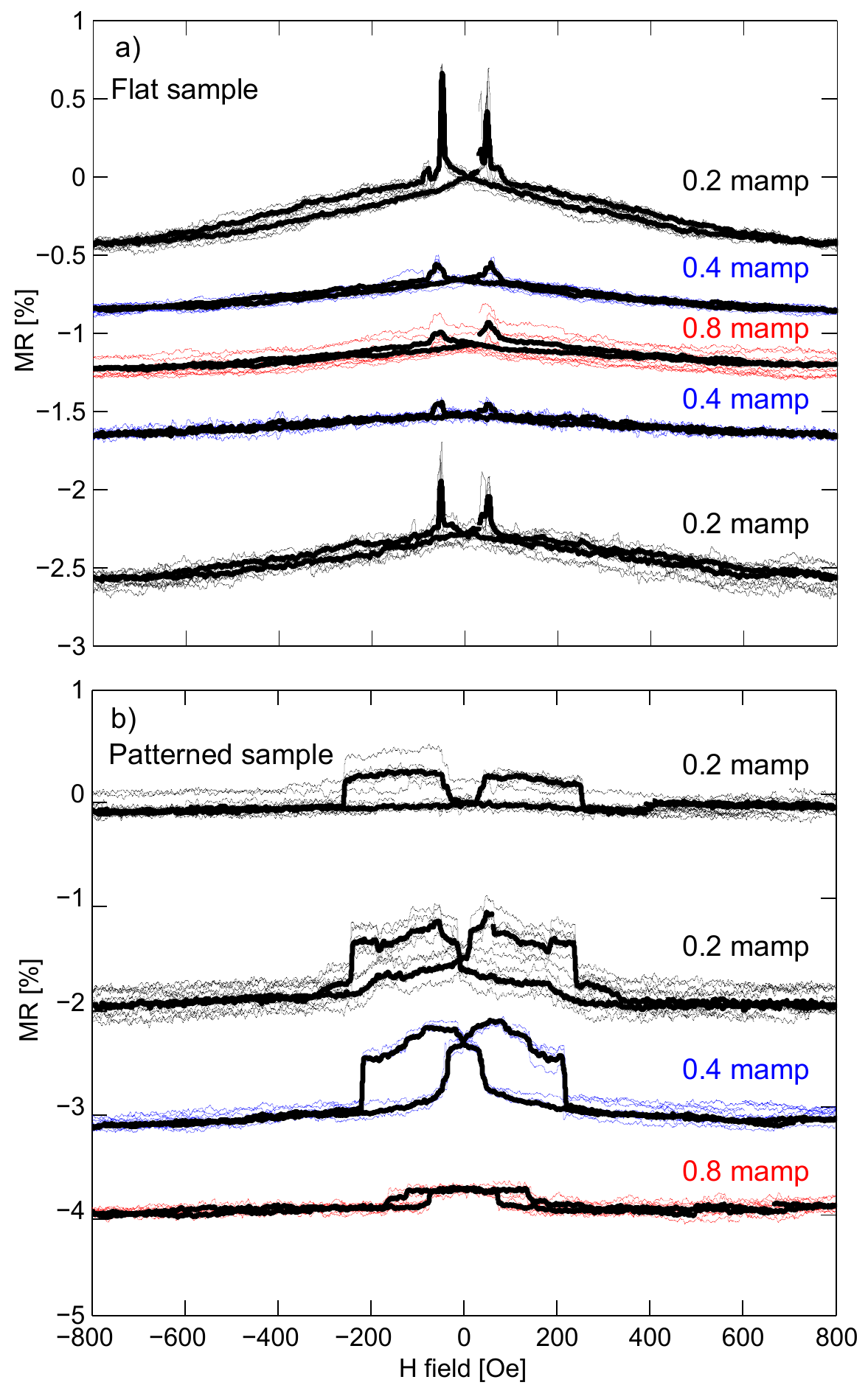}
\caption{\label{fig:MR_3figs} Sharp MR switching signatures from a a) flat  and a b) laterally patterned film at 150 K for various currents, illustrating the difference in coercive fields and current effects from confined geometries. The chronology of the measurements for both panels is from  bottom to the top. }
\end{center}
\end{figure}

The sharp switching with a variable hysteresis depending on point contact (bottom plot figure \ref{fig:MR_examples} and \ref{fig:MR_3figs} a) is only observed in the low amplitude MR sweeps is in accordance with a boundary model interpretation. If the tip happens to be placed particularly close to a wall or boundary and the domains on either side are free to flip regardless of the other, sharp transitions occur.  Both grain boundaries and domain walls  are known to yield such signatures \cite{hwang1996, arnal2007, mathur1999}.  In the absence of sharp switching the low field MR is still present,  which we interpret as an averaged and reduced MR originating from similar processes further away from the tip.

To further elucidate the GB/DW mechanism we compare the sharp MR switching response from a flat (a) and a laterally patterned (b) film at varying bias currents in figure \ref{fig:MR_3figs}. As opposed to the continuous films, the patterned samples do not display the same background low field MR as the rest of the samples. This supports the proposed  probing of magnetic configuration depending on proximity to tip placement; for the patterned samples the situation  is expected to be more binary; chances are that either there is a scattering boundary within the confinement or it is not. 
The crystalline macro-domains in LSMO, which separate the orthogonal twinning orientations \cite{lebedev2001b, maurice2003}, are of similar size ($\sim$ 500 nm) as the structures in our patterned samples and thus consistent with our observations. 

The field dependence of the bulk magnetization is normally reflected in MR curves; in both polycrystalline samples \cite{li1997, xiao1992} and TMR junctions \cite{sun1998}, the coercive field will coincide with switching fields in the MR data. In our samples, however, it is clear that the MR data obtained with the STM tip does not follow the low field bulk magnetization reversal processes presented  in figure \ref{fig:SQUID}, but is shifted to larger fields. We attribute this discrepancy to the locality of our probing;  as seen in the tail of our 150 K M-H curves (figure \ref{fig:SQUID}  b), full saturation is not reached until $\sim$ 100 Oe and could 
encompass the regions probed in our MR measurements.

It is also clear that the patterned samples display a significantly wider field range of high resistance than the flat samples (figure \ref{fig:MR_3figs}). This can be assigned to the induced shape anisotropy and effects from edge roughness in the patterned samples \cite{Blundell2001}.  Similar observations are reported for patterned TMR junctions \cite{sun1998}.

The lateral confinement is also visible through the current dependence of the MR (figure \ref{fig:MR_3figs}). Whereas flat films normally yield stable and reversible field dependence with current bias, the switching fields in patterned films are susceptible to the probe current, indicating a spin transfer torque effect \cite{berger1996, slon1996}. Such torque alters the energetics of nucleation and motion of domain walls during the switching process, and is manifested as perturbations to the coercive fields. The discrepancy in behavior between flat and patterned films in this respect, we attribute to confinement of current in the latter. Typical current densities through our patterned samples reach $\sim$ 10$^6$ amp/cm$^2$, which are  comparable to what is necessary to achieve such  transfer  \cite{slon1996}.

\subsubsection*{High amplitude MR}

We link the high MR observed in the reverse biased diodic sample to transport processes across the measured LSMO/Nb:STO Schottky barrier. In forward bias above the turn-on voltage (true for all datapoints acquired for this polarity), the Schottky barrier is no longer current limiting. We are thus bypassing the barrier, and probing the field dependence of the bulk LSMO transport just as in the non-diodic n$>$0, explaining the identical MR behavior in forward bias for all samples independent of rectifying properties.

However, in opposite voltage polarity (figure \ref{fig:MR_examples} and \ref{fig:MR_norm_summary}), the Schottky barrier is still active, and conduction is mediated by a combination of defect assisted conductance and tunneling. Accompanying the high MR is also a significant magnetic hysteresis,  which we attribute to a magnetically altered interface layer known to exist in the manganites \cite{borges2001, sun1999, angeloni2004}, and the field shift of ~170 Oe fits well with recent reports \cite{huijben2008, kim2010, monsenfmr}. 

The magnetic order in these altered layers is also reduced at elevated temperatures as compared to bulk LSMO behavior, providing an interpretation of the reduced  MR at 300 K. Accordingly, we attribute the high amplitude MR and accompanying hysteresis to the interface region. Control of the interface layer and possible defects is thus a key parameter in order to achieve MR in a reverse biased junction.

The exact mechanism for the MR in this region remains unclear, but it is evident that Schottky barrier is a necessity for the evolution of this MR, altering the transport channels as well as potential and electron density at the diode interface. It has recently been shown that magnetic systems can be susceptible electric fields; through ferromagnetic/ferroelectric interface studies  the use of ferroelectric barriers have enabled switchable control of spin polarization \cite{garcia2010, zhuravlev2005} and even polarization reversal \cite{pantel2012} in LSMO TMR junctions.

\section{Summary}

Local charge transport has been investigated for flat and lithographically patterned (LSMO)$_{40}$/(LMO)$_n$/(LSMO)$_{40}$ heterostructures ($0\leq $ n $ \leq 8$) on Nb:STO substrates. STM based point-contact measurements were used to probe local current-voltage and low field magnetotransport properties. 

Depending on sample type we observe different contributions from the Schottky junction forming at the LSMO/Nb:STO interface; being almost perfect diodic when the current is localized to laterally defined lithographic patterns, while nearly symmetric when probing flat films with an LMO layer inserted. This is attributed to defects in the interface layer in combination with the increased effective diode area arising from scattering of the local current by the LMO inter-layers.

To the low field MR we locate two separate contributions, one originating from the LSMO thin film and the other from the diode region, as deduced through analysis of the I-V, AC and DC characteristics. The thin film contribution is heavily dependent on tip placement, which in our local probe configuration can yield sharp transitions. We argue that this stems from a proximity dependence on distance from GBs or DWs. In laterally patterned samples we observe current dependence on switching fields, that we attribute to the spin torque effect.

The MR contribution from the interface layer in reverse bias is always larger and significant (1.5 \% in at 150 K) with a magnetic coercivity coinciding  with reports of altered interface layers, and is robust over a large range of reverse voltage. 

These results are important for the detailed understanding of MR contributions in complex manganite heterostructures and suggest that device area, interface film properties, defect concentration and nature as well as the geometry, can influence the MR within the same system.

\section{Acknowledgments}

This work was supported by the Norwegian Research Council (NFR) (project number 182037 and 171332/V30) and the Swedish Foundation for International Cooperation in Research and Higher Education (STINT).
The authors would like to thank the MC2 access program and technical crew at Chalmers for cleanroom training and helpful discussions. \\\\

\addcontentsline{toc}{chapter}{Bibliography} 
\bibliographystyle{apsrev}
%


\end{document}